\documentstyle[11pt]{article}

\begin{document}
\title {THE PATH INTEGRAL QUANTIZATION AND THE CONSTRUCTION OF THE S-MATRIX 
OPERATOR IN THE ABELIAN AND NON-ABELIAN CHERN-SIMONS THEORIES}
\author{V.Ya.Fainberg\thanks{Permanent address : P.N.Lebedev Institute of 
Physics, Moscow}$~$\thanks{Supported in part by funds from the RFFI grant no. 
93-02-3379}, N.K.Pak and M.S.Shikakhwa\\
Department of Physics-Middle East Technical University\\
06531 Ankara-Turkey}
\maketitle
\begin {abstract}
 The covariant path integral quantization of the theory of the scalar and 
spinor particles interacting through the Abelian and non-Abelian pure 
Chern-Simons gauge fields is carried out and is shown to be mathematically ill
defined due to the absence of the transverse components of these gauge fields.
 This is remedied by the introduction of the Maxwell or the Maxwell-type (in 
th non-abelian case) term which makes the theory superrenormalizable and 
guarantees its gauge-invariant regularization and renormalization .
 The generating functionals are constructed and shown to be formally the 
same as those of QED (or QCD) 
in 2+1 dimensions with the substitution of the Chern-Simons propagator 
for the photon (gluon) propagator. By constructin the propagator in the 
general case; the existence of two limits; pure 
Chern-Simons and QED (QCD) after renormalization is demonstrated.\\
 By carrying out carefully the path integral quantization of the non-Abelian
 Chern-Simons theories using the De Witt-Fadeev-Popov and the Batalin-Fradkin
-Vilkovisky methods it is demonstrated that there is no need to quantize the 
dimensionless charge of the theory. The main reason is that the action in the 
exponent of the path integral is BRST-invariant which acquires a
zero winding number and guarantees the BRST renormalizability of the model.\\
 The S-matrix operator is constructed, and 
 starting from this S-matrix operator novel topological 
unitarity identities are derived that demand the vanishing of the 
gauge-invariant sum of the imaginary parts of the Feynman diagrams with a 
given number of intermediate on-shell topological photon lines in each order 
of perturbation theory. These identities are illustrated by an explicit 
example.

\end {abstract}
\newpage
\section{Introduction}
$~~~$The past fifteen years witnessed an increasing interest in the theories of
 matter coupled  Chern-Simons (CS)gauge field theories in 2+1 dimensions. From
 one 
point of view, the Euclidean version of such theories can be viewed as giving 
the high temperature bahaviour of 3+1 dimensional models [1]. On the other
hand, in the pioneering works [2,3] it has been shown that the introduction of 
the ( P and T odd ) CS term into the Lagrangian of 2+1 dimensional QED and QCD,
 leads to a very peculiar property : the gauge field splits into two parts; a 
massive part (that acquires a mass in a gauge-invariant manner), and a massless
 part which does not contribute to the free classical Hamiltonian, but leads 
to an additional interaction among the particles. This interaction appears 
also in pure CS theories [4].\\
 In the work [3], it was argued that in the
non-Abelian version of CS theories, the dimensionless combination of the charge
 and the stochastic parameter should be quantized. It was also shown that the 
mass term provides an infrared cut-off in special covariant gauges that 
renders the theory superrenormalizable. \\
 Many works were devoted to the consideration of the one-loop radiative 
corrections to the charge and the stochastic parameter in both the Abelian [5]
and the non-Abelian [6] theories, and a theorem [7] was set which states that 
under very general conditions, there are no further radiative corrections 
beyond the finite one-loop for these parameters.\\
 $~~$An additional thrust into the interest in CS theories was provided by the
interesting results in the non-relativistic domain; essentially the idea of 
Wilczeck that non-relativistic charged particles coupled to pure CS field can
be considered as a phenomenological approach for the description of the "bound
states " of two particles called anyons [8]. This idea found wide acceptance,
and many attempts to apply it in many interesting 
condensed matter phenomena, such as the fractional quantum Hall effect, and 
high temperature superconductivity were made ( see the reviews [9] and the 
references therein ). CS theories  also found applications in the 
field-theoretic formulation of the Aharonov-Bohm effect [10,11].\\
 One of the issues that received considerable interest during the past period 
was the 
canonical quantization of the CS models [2,3,12]. However, some interesting 
points
 like the canonical quantization in a Lorentz covariant gauge still need 
further investigation. Path integral quantization was also considered first
-up to our knowledge- in the works [11,13] where the  generating functional 
was also constructed.\\
 Another issue that did not receive much attention is the following: 
  The free transverse topological photons of the pure CS theory
are absent, while the gauge field propagator is present, and gives significant
contribution to the interaction among the particles. This issue was addressed 
in the work [14], and the so called topological unitarity identities were 
derived. Moreover, the issue of the quantization of the charge in non-abelian
CS theories was not discussed thoroughly beyond the discussion in the works
 [2,3].
 We address this point in the present work.\\
$~~$This paper is a further development of the series of works [11,13,14]. The 
main
goals are, to carry out the path integral quantization and construct 
the generating functional for a wide class of models involving  both the 
Abelian and the non-Abelian CS fields (part II), to construct the 
S-matrix 
operator, and to develop the Feynman rules and formulate a Wick-type theorems
for the CS field (part III), and to illustrate in details the topological
unitarity identities in general, and through a specific example ( part IV). 
Part V is devoted to concluding remarks.   
     
\section{Path Integral Quantization and the Generating Functional}
 The aim of this part is to develop the path integral quantization, and to 
construct the generating functional of the theory of scalar and spinor fields 
interacting through the Abelian and non-Abelian CS field in 2+1 dimensions. 
This can be done through two different approaches : The De Witt-Fadeev-Popov 
(DFP)[15] approach, or the Batalin-Fradkin-Vilkovisky (BFV) approach [16]. The 
latter was developed to quantize gauge theories with both classes of 
constraints and with arbitrary constraint algebra. In our case both approaches
lead to the same result . This is a consequence of the fact that the first
 class 
constraints, both in the Abelian and non-Abelian cases, form a closed algebra,
 and that the structure functions in the algebra of the first class constraints
 are just constants, as will be demonstrated later. Therefore, we shall carry 
out the path integral quantization through the simpler DFP approach, and will 
prove the 
equivalence of both approaches by invoking the latter in the quantization of 
the theory of spinors interacting through the non-abelian CS gauge field. This 
proof is very helpful in understanding the connection between the usual 
canonical quantization and the BFV quantization schemes, and in the 
demonstration of the appearence of the BRST operators of the theory.
\subsection{De Witt-Fadeev-Popov method}
 Scalar particles:\\
 We begin with the theory of charged scalar particles interacting through the 
CS gauge field with the action of this gauge field given slolely by the abelian
 CS term (pure CS field). Following the DFP method, we get for the generating
 functional in the covariant $\alpha$-gauge the expression [11,13] :
\begin{eqnarray}
Z[J_{\mu},j,j^*]&=&Z_0^{-1}\int DA_{\mu}(x)d\varphi^*(x)D\varphi(x)\exp
\{iS_{CS}+iS_g+iS_m\nonumber\\
&+&i\int d^3x(J_{\mu}(x)A^{\mu}(x)+j^*(x)\varphi(x)+j(x)\varphi^*(x))\}
\end{eqnarray}
where
\begin{eqnarray}
Z_0&=&Z(0,0,0,)\\
S_{CS}&=&{\mu\over 2}\int d^3x \varepsilon_{\mu\nu\lambda}A^{\mu}(x)
\partial^{\nu}A^{\lambda}(x)\\
S_g&=&{-1\over 2\alpha}\int (\partial_{\mu}A^{\mu})^2d^3x\\
S_m&=&\int d^3x\left(\varphi^*(x)(D_{\mu}D^{\mu}-m^2)\varphi(x)-
\lambda(\varphi^*(x)\varphi(x))^2\right)
\end{eqnarray}
 Here, $J_{\mu}(x),~j(x)$ and $j^*(x)$ are external sources, $e$ and $m$ are 
respectively the charge and the mass of the scalar field, and $D_{\mu}=
(\partial_{\mu}-ieA_{\mu})$. The metric is taken as $g_{\mu\nu}=diag(1,-1,-1)$
. The Greens functions of the theory are defined as usual by varying the above
 generating functional, eq.(1) with respect to the sources.
 For example, the free propagator of the CS field is defined as:
\begin{eqnarray}
D_{\mu\nu}(x-x')=(-i)^2{\delta^2\over \delta J_{\mu}(x)J_{\nu}(x')}
Z[J_{\mu},j,j^*]\left.\right|_{J_{\mu}=j=j^*=e=0}
\end{eqnarray}
However, the functional (1) has an essential defect in that the path integral 
over the gauge field is not mathematically well-defined . This is 
because in pure CS theory, there are no transverse components of the gauge 
field. The natural way to overcome this difficulty is to introduce into the 
total action in the exponent of the path integral (1) the Maxwell term
\begin{eqnarray}
S_M={-1\over 4\gamma}\int d^3xF_{\mu\nu}(x)F^{\mu\nu}(x)
\end{eqnarray}
 Such a term is the only gauge-invariant bilinear term in $A_{\mu}$ that 
guarantees gauge-invariant regularization and renormalization of the theory.
This term, not only leads to the convergence of the path integral over $
A_{\mu}$, but also plays the role of a regularization factor since the 
resulting theory becomes superrenormalizable [2,3].\\
 It is necessary here to make some important remarks on the dimensions of the 
parameters and the fields of the theory. We have some arbitrariness in the 
choice of the dimensions of the statistical parameter $\mu$, the charge $e$ and
 the factor $\gamma$ in eqs.(3),(5) and (7). However, if we require the 2+1 
dimensional
 matter-coupled CS theory to have some relation with the real world, and 
so that it 
arises after compactification on the $\sim{1\over\gamma}$ layer of QED in 3+1 
dimensions [17]with the parity violating term ${\mu\over 4}\int F_{\mu\nu}
\tilde
F^{\mu\nu}d^4x$ where $\tilde F^{\mu\nu}={1\over 2}\varepsilon^{\mu\nu\lambda
\sigma} F_{\lambda\sigma}$ then the charge $e$ and the parameter $\mu$ are to 
be chosen dimensionless, whereas $[A_{\mu}]= x^{-1}~,~[\varphi]=
 x^{{-1\over 2}}$ and $[\gamma]= x^{-1}$. In the following, we will 
adopt this convention of the dimensions \footnote{If one makes the change of 
variables $A_{\mu}\to A_{\mu}'={A_{\mu}\over\sqrt{\gamma}}~,~e\to e'=e\sqrt
{\gamma}~,~\mu\to\mu'={\mu\over\gamma}$ then one gets the conventions used in 
the works [2,3]}. So, after introducing the Maxwell term the generating 
functional takes the form
\begin{eqnarray}
Z[J_{\mu},j,j^*]&=&Z_0^{-1}\int DA_{\mu}(x)D\varphi^*(x)D\varphi(x)
\exp\{i(S_{CS}+S_M+S_g+S_m)\nonumber\\
&+&i\int d^3x(J_{\mu}(x)A^{\mu}(x)+j^*(x)\varphi(x)+j(x)\varphi^*(x))\}
\end{eqnarray}
 This can be  formally written in the alternative form 
\begin{eqnarray}
Z[J_{\mu},j^*,j]=Z_0^{-1}\int D\varphi^*(x)D\varphi(x)\exp(ie^2\int d^3x
{\delta^2\over \delta J_{\mu}(x)\delta J^{\mu}(x)})\nonumber\\
\times\int DA_{\mu}(x)\exp\{i(S_{CS}+S_M+S_g+\tilde S_m)\nonumber\\
+i\int d^3x(J_{\mu}(x)A^{\mu}(x)+j^*(x)\varphi(x)+\varphi^*(x)j(x))\}
\end{eqnarray}
where $\tilde S_m$ does not contain the term $e^2A_{\mu}A^{\mu}$ in eq.(5),i.e
\begin{eqnarray}
\tilde S_m=-\int d^3x(ieA_{\mu}(x)(\varphi^*(x)\partial_{\mu}\varphi(x)
-\varphi(x)\partial_{\mu}\varphi^*(x))+\lambda(\varphi^*(x)\varphi(x))^2)
\end{eqnarray}
After integrating over $A_{\mu}$ in eq.(9) we get :
\begin{eqnarray}
Z[J_{\mu},j,j^*]=Z_0^{-1}\int D\varphi^*(x)D\varphi(x)\exp\{ie^2\int d^3x
\varphi^*(x)\varphi(x){\delta^2\over\delta J_{\nu}(x)\delta J^{\nu}(x)}\}
\nonumber\\
\exp\{{i\over 2}\int d^3xd^3yI_{\mu}(x)D^{\mu\nu}(x-y)I_{\nu}(y)-\lambda\int 
d^3x(\varphi^*(x)\varphi(x))^2\nonumber\\
+i\int d^3x(j^*(x)\varphi(x)+j(x)\varphi^*(x))\}
\end{eqnarray}
where
\begin{eqnarray}
I_{\mu}(x)=J_{\mu}(x)+ie\int d^3x(\varphi^*(x)\partial_{\mu}\varphi(x)-\varphi
(x)\partial_{\mu}\varphi^*(x))
\end{eqnarray}
and $D_{\mu\nu}(x-y)$ is the CS gauge field's Greens function defined by the 
equation:
\begin{eqnarray}
\int d^3x'{\delta^2(S_{CS}+S_g+S_M)\over \delta A^{\mu}(x)\delta A^{\lambda}
(x')}D^{\lambda\nu}(x'-y)=g^{\nu}_{\mu}\delta^3(x-y)
\end{eqnarray}
or,
\begin{eqnarray}
[{1\over\gamma}(\Box_xg_{\mu\lambda}-\partial_{\mu}\partial_{\lambda})+{1\over
\alpha}\partial_{\mu}\partial_{\lambda}+\mu\varepsilon_{\mu\lambda\rho}
\partial_x^{\rho}]D^{\lambda\nu}(x-y)=\delta^3(x-y)g^{\nu}_{\mu}
\end{eqnarray}
 The solution of eq.(14) is [2,3]:
\begin{eqnarray}
D_{\lambda\nu}(x)={1\over (2\pi)^3}\int d^3pe^{ipx}\left[-\gamma
{(g_{\lambda\nu}-{p_{\nu}p_{\lambda})\over p^2}\over (p^2-\gamma^2\mu^2+
i\epsilon)}+{i\varepsilon_{\lambda\nu\rho}
p^{\rho}\over\mu(p^2-\gamma^2\mu^2+i\epsilon)}\right.\nonumber\\
\left.-{i\varepsilon_{\lambda\nu\rho}p^{\rho}\over \mu(p^2+i\epsilon)}
-{\alpha p_{\lambda}p_{\nu}\over (p^2+i\epsilon)^2}\right]
\end{eqnarray}
 We note that the above Greens function (or propagator) consists of two parts:
 The first two terms describe the propagation of a real massive photon with 
mass equal to $\gamma\mu$; the third term describes the propagation of a 
 topological massless photon, and the last term is pure gauge term. The
 appearence of massive photons in a gauge-invariant manner is a well-known 
peculiar property of CS theory, and is independent of coupling to matter fields
 [2,3]. To show that the topological term in eq.(15) does not contribute to 
the 
tensor $F_{\mu\nu}$ of the gauge field, we construct the general solution of 
the classical equations of motion of the field $A_{\mu}$ (eq.(14)). This is 
given as :
\begin{eqnarray}
A_{\mu}(x)&=&4\pi\int Im D_{\mu\nu}(p){\it e}_{\delta}^{\nu}a^{\delta}(p)e^{ik
x}d^3p\nonumber\\
&=&{1\over 2\pi}\int d^3pe^{ipx}\left[-\gamma(({\it e}^{\delta}_{\mu}(p)
-{p_{\mu}p_{\nu}\over p^2}{\it e}^{\nu}_{\delta}(p))+{i\over\gamma\mu}
\varepsilon_{\mu\nu\rho}p^{\rho}{\it e}^{\nu}_{\delta}(p))\delta(p^2-\mu^2
\gamma^2)\right.\nonumber\\
&-&{i\over\mu}\varepsilon_{\mu\nu\rho}{\it e}^{\nu}_{\delta}(p)p^{\rho}
\delta(p^2)
-({p_{\mu}p_{\nu}\over p^2-\mu^2\gamma^2}){\it e}^{\nu}_{\delta}(p)\delta(p^2)
\nonumber\\
&+&{\alpha\over 2}p_{\mu}\left({\partial\over\partial p^{\nu}}\delta(p^2)
\right) {\it e}^{\nu}_{\delta}(p)\left.\right]a^{\delta}(p)
\end{eqnarray}
Here, $ImD_{\mu\nu}(p)$ is the imaginary part of the propagator $D_{\mu\nu}$ in
 eq.(15) in the momentum space representation; $e^{\nu}_{\delta}(p), \delta=0,1
,2,$ are three mutually orthogonal polarization vectors which satisfy
$p_{\mu}e^{\mu}_{\delta}(p)=0$. This choice corresponds to the gauge 
$\partial_{\mu} A^{\mu}=0$. In the general case, the free solution $A_{\mu}(x)
$ in eq.(16) represents the sum of two independent parts : The terms 
proportional to $\delta(p ^2-\mu^2\gamma^2) $ correspond to a real massive 
photon which contributes to the free Hamiltonian; the fourth and fifth terms 
are the topological parts of the gauge field which do not contribute to the 
classical free
Hamiltonian, but give non-trivial contribution to the propagator (see eq.(15))
, and the last term is merely a gauge term that can be removed by a gauge 
transformation. It is easy to see that the topological part of $A_{\mu}$ does 
not contribute to $F_{\mu\nu}$:
\begin{eqnarray}
F_{\mu\nu}&=&\partial_{\mu}A_{\nu}-\partial_{\nu}A_{\mu}\nonumber\\
&=&{1\over 2\pi\mu}\int d^3pe^{ipx}\delta(p^2)a^{\delta}(p)(p_{\mu}
\varepsilon_{\nu\lambda\rho}-p_{\nu}\varepsilon_{\mu\lambda\rho})
{\it e}^{\lambda}_{\delta}p^{\rho}
\end{eqnarray}
multiplying both sides by $\varepsilon_{\sigma\mu\nu}$ we get
\begin{eqnarray}
\varepsilon^{\sigma\mu\nu}F_{\mu\nu}={1\over \mu\pi}\int d^3pe^{ipx}\delta(p^2)
a^{\delta}(p)({\it e}^{\sigma}_{\delta}(p)p^2-p_{\mu}{\it e}^{\mu}_{\delta}
p^{\sigma})=0
\end{eqnarray}
since
\begin{eqnarray}
p_{\mu}{\it e}^{\mu}_{\alpha}=0
\end{eqnarray}
 As for the the massive part of the solution (16), excluding the second term 
in this equation in view of (19) above, then we have for the massive part
\begin{eqnarray}
A_{\mu}(x)={-1\over 2\pi}\int d^3pe^{ipx}\gamma\left({\it e}^{\delta}_{\mu}(p)
-{i\over\mu\gamma}\varepsilon_{\mu\nu\rho}p^{\rho}{\it e}^{\nu}_{\delta}\right)
a_{\delta}(p)\delta(p^2-\mu^2\gamma^2)
\end{eqnarray}
and this gives a non-vanishing contribution to  $F_{\mu\nu}$. We shall 
return later to the question of quantization of this $A_{\mu}$ in connection 
with the construction of the S-matrix of the theory (see part III).\\
 Returning to the general expression for the Greens function of the gauge field
 , we stress that formally it is possible to consider two limiting procedures
in eq.(15). First, if $\gamma\to\infty$ we obtain:
\begin{eqnarray}
\lim_{\gamma\to\infty}D_{\lambda\nu}=D^{CS}_{\lambda\nu}={-1\over(2\pi)^3}\int 
d^3p
e^{ipx}\left({i\varepsilon_{\nu\lambda\rho}p^{\rho}\over\mu(p^2+i\epsilon)}
+{\alpha p_{\lambda}p_{\nu}\over (p^2+i\epsilon)^2}\right)
\end{eqnarray}
which is just the propagator of the pure CS theory. In the limit $\mu\to 0$, we
 get the usual Feynman propagator in 2+1 dimensional QED for massless photons:
\begin{eqnarray}
\lim_{\mu\to 0}D_{\lambda\nu}(x)=D^M_{\lambda\nu}(x)={-\gamma\over (2\pi)^3}
\int d^3pe^{ipx}{\left(g_{\lambda\nu}-({p_{\lambda}p_{\nu}\over p^2+i\epsilon})
(1-{\alpha\over\gamma})\right)\over(p^2+i\epsilon)}
\end{eqnarray}
 In both cases, we have from eq.(16) $A_{\mu}$ as
\begin{eqnarray}
\lim_{\gamma\to\infty}A_{\mu}(x)&=&{-1\over 2\pi}\int d^3pe^{ipx}\left[\left(
{i\over\mu}\varepsilon_{\mu\nu\rho}p^{\rho}-
{\alpha\over 2}p_{\mu}{\partial\over\partial p^{\nu}}\right)\delta(p^2)\right]
{\it e}^{\nu}_{\delta}(p)a^{\delta}(p)\nonumber\\
&=&A^{CS}_{\mu}
\end{eqnarray}
and
\begin{eqnarray}
\lim_{\mu\to 0}A_{\mu}(x)={-\gamma\over 2 \pi}\int d^3pe^{ipx}\left[\left(
g_{\mu\nu}+{(1-{\alpha\over\gamma})\over 2}p_{\mu}{\partial\over
\partial p^{\nu}}
\right)\delta(p^2)\right]{\it e}^{\nu}_{\delta}(p) a^{\delta}(p)
\end{eqnarray}
 The limits are to be taken after renormalization, 
recalling that in our theory one has only the finite one-loop correction to 
$\mu$ (or $\gamma$) [5-7].

Spinor CS theory:

$~~~$Let us consider now the theory of spinor particles interacting through the
 CS gauge field. The DFP method gives the following expression for the 
generating functional in this case [14]:
\begin{eqnarray}
Z[J_{\mu},\eta,\bar\eta]=Z_0^{-1}\int DA_{\mu}(x)D\bar\psi(x)D\psi(x)\exp\{
iS_{CS}+iS_M+iS_g+iS_{\psi}\nonumber\\
+i\int d^3x(J_{\mu}(x)A^{\mu}(x)+\bar\eta(x)\psi(x)+\bar\psi(x)\eta(x))\}
\end{eqnarray}
where $Z_0=Z(0,0,0)~;~S_{CS},S_g$ and $S_M$ are defined by eqs.(3),(4) and 
(7) respectively, and 
\begin{eqnarray}
S_{\psi}=\int d^3x\bar\psi(x)(iD\!\!\!\!/-m)\psi(x)
\end{eqnarray}
where,
\begin{eqnarray}
D\!\!\!\!/=D_{\mu}\gamma^{\mu}~~~~~~,~~~D_{\mu}=(\partial_{\mu}-ieA_{\mu})
\end{eqnarray}
and the Dirac matrices are defined as 
\begin{eqnarray}
\gamma_0=\sigma_3~~~,~~\gamma_i=i\sigma_i~~~~,i=1,2
\end{eqnarray}
where $\sigma$'s are the Pauli spin matrices. The $\gamma$-matrices satisfy
\begin{eqnarray}
\{\gamma_{\mu},\gamma_{\nu}\}=2g_{\mu\nu}~~~,~~\gamma_{\mu}\gamma_{\nu}=
g_{\mu\nu}-i\varepsilon_{\mu\nu\lambda}\gamma^{\lambda}
\end{eqnarray}
$\psi(x)$ and $\bar\psi(x)=\psi(x)^{\dagger}\gamma_0$ are the two-component 
Grassmann spinors, $\eta$ and $\bar \eta$ are Grassmann sources. Integrating 
over $A_{\mu}(x)$ in eq.(25) we get :
\begin{eqnarray}
Z[J_{\mu},\bar\eta,\eta]=Z_0^{-1}\int D\bar\psi(x)D\psi(x)\exp\{{i\over 2}
\int d^3xd^3y\tilde I_{\mu}(x)D^{\mu\nu}(x-y)\tilde I_{\nu}(y)\nonumber\\
+i\int d^3x(\bar\eta(x)\psi(x)+\bar\psi(x)\eta(x))\}
\end{eqnarray}
where
\begin{eqnarray}
\tilde I_{\mu}(x)=J_{\mu}(x)+e\bar\psi(x)\gamma_{\mu}\psi(x)
\end{eqnarray}
 and $D_{\mu\nu}(x-y)$ is the bare CS field propagator which is the same as in 
the scalar case, eq.(15). Here also, as in the scalar case, one can consider 
the limits (after renormalization) $\gamma\to\infty$ and $\mu\to 0$ to get the 
propagators of pure CS field and 2+1 dimensional QED respectively.\\
Non-Abelian CS theory:

$~~$The path integral quantization of theories with the non-Abelian CS gauge 
field is a bit more complicated than the Abelian one, so we consider it in some
 detail. We start with the theory of the gauge field without coupling to matter
, i.e CS gluodynamics, defined by the Lagrangian
\begin{eqnarray}
{\cal L}={\cal L}_M+{\cal L}_{CS}
\end{eqnarray}
${\cal L}_M$ is the usual Yang-Mills Lagrangian in 2+1 dimensions,
\begin{eqnarray}
{\cal L}_M&=&{-1\over 2\gamma}tr\left(F_{\mu\nu}(x)F^{\mu\nu}(x)\right)
\nonumber\\
F_{\mu\nu}&=&\partial_{\mu}A_{\nu}(x)-\partial_{\nu}A_{\mu}(x)+g[A_{\mu}(x),
A_{\nu}(x)]
\end{eqnarray}
${\cal L_{CS}}$ is the non-Abelian CS term
\begin{eqnarray}
{\cal L}_{CS}=-\mu\varepsilon^{\mu\nu\lambda}tr(A_{\mu}(x)\partial_{\nu}
A_{\lambda}(x)+{2i\over 3}gA_{\mu}(x)A_{\nu}(x)A_{\lambda}(x))
\end{eqnarray}
 The gauge group is $SU(N)$. In matrix notation
\begin{eqnarray}
A_{\mu}=A_{\mu}^at^a~~~~;~~~~F_{\mu\nu}=F_{\mu\nu}^at^a
\end{eqnarray}
The $t^a$'s are antihermitian matrices in the fundamental representation of 
the group
\begin{eqnarray}
[t^a,t^b]=if^{abc}t_c~~~~,~~~tr(t^at^b)={1\over 2}\delta^{ab}
\end{eqnarray}
$f^{abc}$ are the structure constants of the $SU(N)$ group.\\
 To see the difference of the non-Abelian case from the Abelian one, consider 
a general gauge transformation
\begin{eqnarray}
A_{\mu}(x)\to U^{-1}(A_{\mu}(x)-{i\over g}\partial_{\mu})U
\end{eqnarray}
${\cal L}_M$ is gauge-invariant, ${\cal L}_{CS}$ is not [2,3];
\begin{eqnarray}
\int d^3x{\cal L}_{CS}\to\int d^3x{\cal L}_{CS}-{i\mu\over g}\int d^3x
\varepsilon^{\mu\nu\lambda}\partial_{\mu}tr\left((\partial_{\nu}U)U^{-1}
A_{\lambda}\right)\nonumber\\
+{8\pi^2\mu\over g}iw
\end{eqnarray}
where
\begin{eqnarray}
w={1\over 24g\pi^2}\int d^3x\varepsilon^{\mu\nu\lambda}tr\left[(U^{-1}
\partial_{\mu}U)(U^{-1}\partial_{\nu}U)(U^{-1}\partial_{\lambda}U)\right]
\end{eqnarray}
 If we suppose that at $||x||=\sqrt{x_0^2+\vec x^2}\to\infty~,~A_{\mu}\to 0$ 
faster than ${1\over ||x||}$ then the second term in (38) vanishes. The last 
term
, however, coincides in {\it euclidean\/} space, with the so called homotopy 
class 
 or winding number, and is equal to $0,\pm 1,\pm 2,...$ . This result follows 
from the fact that if 
\begin{eqnarray}
U(x)_{||x||\to\infty}\to 1,
\end{eqnarray}
then 3-dimensional space can be mapped onto $S_3$; for $SU(2)$ group $U(x)$ 
realizes the mapping $S_3\to S_3$ and the winding number is equal to the 
degree of mapping $S_3$ to the $SU(2)$ group. On the classical level, the 
gauge-transformation (37) results in 
\begin{eqnarray}
S_{CS}\to S_{CS}+constant
\end{eqnarray}
 It is clear that this constant does not influence the equations of motion or 
any physical quantity.\\
 Now, we use the Fadeev-Popov trick to quantize the theory. Formally, the 
vacuum functional of the theory is
\begin{eqnarray}
Z_0=N\int DA_{\mu}\exp i\{S_M+S_{CS}\}
\end{eqnarray}
where $N$ is a normalization factor that will be defined later. Introducing 
into the formal equation (42) the identity operator in a general Lorentz 
covariant gauge
\begin{eqnarray}
I=\bigtriangleup(A)\int D\mu(G)\delta(\partial^{\mu}A_{\mu}^G-f(x))
\end{eqnarray}
where $D\mu(G)$ is the measure of the $SU(N)$ group, and 
\begin{eqnarray}
A_{\mu}^G=U^{-1}(A_{\mu}-{i\over g}\partial_{\mu})U~~~~U\in G.
\end{eqnarray}
Eq.(43) defines the Fadeev-Popov determinant $\bigtriangleup (A)$.\\
 We know that in perturbation theory we can forget about the Gribov ambiguity 
[18] and consider only contributions to the functional integral from elements 
near the identity of the group $G$;
\begin{eqnarray}
U=1+i\lambda(x)+O(\lambda^2)~~~~~;~~~~\lambda=\lambda^at^a
\end{eqnarray}
 where $\lambda^a(x)$ is infinitesimally small for all x. This means that in 
DFP method we 
 must consider gauge transformations which belongs to the zero homotopy class 
for which $w=0$
 since $\lambda(x)$ must go to zero when $||x||\to\infty$ \footnote{In the 
general case when $U=e^{i\tau^a\lambda^a(x)}$ and $||x||\to\infty~,~\lambda(x)
=\sqrt{(\lambda^a)^2}\to 2\pi n$ where $n$ is the winding number.}. 
Substituting the identity operator
(43) into the expression (42), we get after the conventional manipulations
\begin{eqnarray}
Z_0=N\Omega(G)\int DA_{\mu}(x)D\bar{\cal C}(x)D{\cal C}(x)\exp\{i(S_M+S_{CS}
+S_g)\}
\end{eqnarray}
Here $\Omega(G)$ is the infinite group volume, and 
\begin{eqnarray}
S_g=\int d^3xtr\left({-1\over 2\alpha}(\partial_{\mu}A^{\mu}(x))^2+
\partial_{\mu}\bar{\cal C}^a(x)(D^{\mu ab}{\cal C}^b(x))\right)
\end{eqnarray}
where ${\cal C}(x)$ and $\bar {\cal C}(x)$ are the well-known Fadeev-Popov
 ghosts that are 
scalar Grassmann fields, and 
\begin{eqnarray}
D_{\mu}^{ab}=\partial_{\mu}\delta^{ab}+gf^{abc}A_{\mu}^c(x)
\end{eqnarray}
Thus, the generating functional of the theory is now given by the expression:
\begin{eqnarray}
Z[J_{\mu},\eta,\bar\eta]=Z^{-1}(0,0,0)\int DA_{\mu}(x)D\bar{\cal C}(x)D{\cal C}
(x)\exp\{i(S_M+S_{CS}+S_g)\nonumber\\
+i\int d^3x(J_{\mu}^a(x)A^{\mu}_a(x)+\bar\eta^a(x){\cal C}^a(x)+\bar{\cal C}^a
(x)\eta^a(x))\}
\end{eqnarray}
here,
\begin{eqnarray}
Z(0,0,0)=Z(J_{\mu},\bar\eta,\eta)\left.\right|_{J_{\mu}=\bar\eta=\eta=0}
\end{eqnarray}
 With the above generating functional, the expectation value of any observable
 is well-defined. For example
\begin{eqnarray}
\left<\hat{\cal O}(\hat A,\hat{\bar{\cal C}},\hat{\cal C})\right>&=&
{\cal O}({-i\delta\over\delta J^a_{\mu}(x)},{-i\delta^l\over\delta\eta^b(x)},
{i\delta^l\over\delta\eta^c(x)})Z[J_{\mu},\bar\eta,\eta]\left.\right|_{J_{\mu}=
\bar\eta=\eta=0}\nonumber\\
&=&Z^{-1}(0,0,0)\int DA_{\mu}D\bar{\cal C}(x)D{\cal C}(x)
{\cal O}(A,\bar{\cal C},{\cal C})\nonumber\\
&\times&\exp\{i(S_M+S_{CS}+S_g)\}
\end{eqnarray}
This expression does not change under any gauge transformation of the total 
action .\\
 $~~~$ We have seen above that the exponent in the generating functional of the
 theory contains after quantization the term $S_g$ that violates 
gauge-invariance.
However, the total action in the exponent preserves invariance under a special 
class of BRST [19] gauge supertransformations
\begin{eqnarray}
A_{\mu}^a(x)\to A_{\mu}^a(x)+(D_{\mu}{\cal C}(x))^a\epsilon\\
{\cal C}^a(x)\to{\cal C}^a(x)-{1\over 2}f^{abc}{\cal C}^b(x){\cal C}^d(x)
\epsilon\\
\bar{\cal C}^a(x)\to\bar{\cal C}^a(x)+{1\over\alpha}(\partial_{\mu}A^{\mu a}(x)
)\epsilon
\end{eqnarray}
where $\epsilon$ is an $x$-independent Grassmann parameter $(\epsilon^2=0)$;
\begin{eqnarray}
\{\epsilon,\bar{\cal C}\}_+=\{\epsilon,{\cal C}\}_+=0=\{\epsilon,A_{\mu}(x)\}_-
\end{eqnarray}
It is well-known that $S_g$ does not change under these transformations. If we 
 formally write down the transformation law of $A_{\mu}(x)$ in the form
\begin{eqnarray}
A_{\mu}'=U^{-1}(A_{\mu}-{i\over g}\partial_{\mu})U
\end{eqnarray}
where
\begin{eqnarray}
U=\exp\{it^a{\cal C}^a\epsilon\}=1+it^a{\cal C}^a\epsilon
\end{eqnarray}
then under this transformation
\begin{eqnarray}
S_{CS}\to S_{CS}+{8\pi^2\mu^2\over g^2}iw=S_{CS}
\end{eqnarray}
since $w=0$ because $\epsilon^2=0$.\\
 It is very important to stress that the BRST-invariance of the CS theory 
ensures satisfying all the Ward-Fradkin-Takahishi-Slavnov-Taylor identities 
[20], and therefore the gauge-invariant renormalizability of the theory [21].
  We thus come to the conclusion that in the framework of perturbation
 theory, it is not necessary to quantize the charge in CS gluodynamics. The
 same holds also if coupling to matter is introduced into the theory as well.\\
$~~$If one introduces spinor field into the theory (CS quantum chromodynamics 
(CSQCD)), then it is straight forward to generalize the generating functional 
eq.(49) to this case. The resulting expression is
\begin{eqnarray}
Z[J_{\mu}^a,\bar\eta,\eta]=Z_0^{-1}\prod_a\int DA_{\mu}^a(x)D\bar\psi(x)
D\psi(x)\exp\{i(S_{CS}+S_M+S_g+\tilde S_{\psi})\nonumber\\
+i\int d^3x(J_{\mu}^a(x)A^{\mu}_a(x)+\bar\eta(x)\psi(x)+\bar\psi(x)\eta(x))\}
\end{eqnarray}
Here $S_{CS},S_M$ and $S_g$ were defined earlier,eqs.(3),(4) and (7),and 
\begin{eqnarray}
\tilde S_{\psi}=\int d^3x\bar\psi_i(x)(\partial\!\!\!/+eA\!\!\!/(x)-m)_{ij}
\psi_j(x)
\end{eqnarray}
$i,j=1,...,N$ above are the color indices of  the $SU(N)$ group in the 
fundamental 
representation.It is straight forward to write down the Feynman propagator of
the non-Abelian gauge field; it will differ from 
the abelian one only by the appearence of color indices viz.
\begin{eqnarray}
D_{\mu\nu}^{ab}(x)=\delta^{ab}D_{\mu\nu}(x)
\end{eqnarray} 
$~~~$Before leaving this subsection, we would like to emphasize that starting 
from the generating functionals for the various models that have been 
considered so far, one can construct all the propagators and the primitive 
vertices, and thus develop the Feynman rules for perturbation theory. For 
example, the Feynman propagator for the scalar field is given as
\begin{eqnarray}
G(x-y)&=&i{\delta^2Z[J_{\mu},j,j^*]\over\delta j(x)\delta j^*(y)}\left.\right|_
{J_{\mu}=j=j^*=e=0}\nonumber\\
&=&{1\over (2\pi)^3}\int d^3p{e^{ip(x-y)}\over p^2-m^2+i\epsilon}
\end{eqnarray}
Similarly, we have for the spinor propagator from eq.(59)
\begin{eqnarray}
S(x-y)&=&(-i)^2{\delta^l\delta^rZ\over\bar\eta(x)\eta(y)}\left.\right|_{J_
{\mu}=\eta=\bar\eta=e=0}\nonumber\\
&=&{1\over (2\pi)^3}\int d^3p{e^{ip(x-y)}\over p\!\!\!/-m}
\end{eqnarray} 
\subsection{Path Integral Quantization Of Pure Chern-Simons Quantum 
Chromodynamic by Batalin-Fradkin-Vilkovisky
Method}
Here, we shall show how to construct the generating functional of the theory 
of spinors coupled to the non-Abelian CS field by the BFV method. We shall 
consider however, a theory where the gauge field kinetic action is given solely
 by the non-abelian CS term, i.e pure CSQCD. This will have a more complicated 
constraint structure than the one with the Maxwell term (eq.(7)) included.The 
BFV quantization makes  the BRST symmetry of the theory, that 
is generated by the operator $\Omega$ introduced below, more transparent. We 
start with the classical action:
\begin{eqnarray}
S=S_{CS}+S_{\psi}
\end{eqnarray}
where $S_{CS}$ and $S_{\psi}$ are given by eqs.(3) and (26). The action 
can be written in a more transparent form:
\begin{eqnarray}
S_{CS}&=&-{\mu\over 2}\int d^3x(A_0^a(x)\varepsilon_{ij}F^{ija}+
\varepsilon_{ij}\dot A^{ia}(x)A^{ja}(x)+{g\over 3}f^{abc}
\varepsilon_{\mu\nu\lambda}A^{\mu}_a(x)A^{\nu}_b(x)A^{\lambda}_c(x))\\
S_{\psi}&=&\int d^3x\left( \bar\psi(x)(i\gamma_0\partial_0-
i\vec\gamma.\vec\nabla -m)\psi(x)-gA_{\mu}(x)\bar\psi(x)\gamma^{\mu}
\psi(x)\right)
\end{eqnarray}
The canonical momenta of the theory turn out to be all primary constraints:
\begin{eqnarray}
\pi_i^a&=&{\delta{\cal L}\over\delta\dot A^{ia}}={-\mu\over 2}
\varepsilon_{ij}
A^{ja}~~~~~;~~\theta_i^a\equiv \pi_i^a+{\mu\over 2}\varepsilon_{ij}A^{ja}
\approx 0\\
\pi_{\psi}&=&{\delta^r{\cal L}\over\delta\dot\psi}=i\psi^{\dagger}~~~~~~~~~~~~
~~;~~
\theta_3\equiv\pi_{\psi}-i\psi^{\dagger}\approx 0\\
\pi_{\psi^{\dagger}}&=&{\delta^l{\cal L}\over\delta\dot\psi^{\dagger}}=0
~~~~~~~~~~~~~~;~~\theta_4\equiv\pi_{\psi^{\dagger}}\approx 0\\
\pi_0^a&=&{\delta{\cal L}\over\delta\dot A_0^a}=0~~~~~~~~~~~~~~ ;~~
G^a\equiv\pi_0^a\approx 0
\end{eqnarray}
 The standard Poisson brackets are:
\begin{eqnarray}
\{\psi(x),\pi_{\psi}(y)\}&=&\{\psi^{\dagger}(x),\pi_{\psi^{\dagger}}(y)\}
=\delta(\vec x-\vec y)\\
\{A_{\mu}^a(x),\pi_{\nu}^b(y)\}&=&g_{\mu\nu}\delta^{ab}\delta(\vec x-\vec y)
\end{eqnarray}
 $\theta_i,\theta_3$ and $\theta_4$ are second class constraints, while $G^a$ 
is first class. The presence of the second class constraints motivates one to 
define the Dirac brackets [22] using these constraints. These can be worked 
out easily, and the ones that differ from the Poisson bracket are:
\begin{eqnarray}
\{\psi(x),\psi^{\dagger}(y)\}_D&=&i\delta(\vec x-\vec y)\\
\{A_i^a(x),A_j^b(y)\}_D&=&{-1\over\mu}\delta^{ab}\varepsilon_{ij}\delta(\vec x
-\vec y)\\
\{A_i^a(x),\pi_j^b(y)\}_D&=&{1\over 2}g_{ij}\delta^{ab}\delta(\vec x-\vec y)
\end{eqnarray}
The Hamiltonian assumes the form
\begin{eqnarray}
{\cal H}&=&{\cal H}_0+A_0^aT^a\nonumber\\
&=&\bar\psi(x)(i\vec\gamma.\vec\nabla+m)\psi(x)-\vec A(x).\vec J(x)\nonumber\\
&+&A_0^a(x)\left(\right.
J_0^a(x)+{\mu\over 2}\varepsilon_{ij}F^{ija}
+g{\mu\over 2}f^{abc}\varepsilon
_{ij}A^{ib}(x)A^{jc}(x)\left.\right)
\end{eqnarray}
${\cal H}_0$ is the Hamiltonian on the constraint surface, and $T^a$ is a 
 first class constraint that is the analogue of Gauss' law constraint in QCD 
, and here also it is the generator of the gauge symmetry.
$A_0$ appears here, as is the case in QED and QCD, as a Lagrange multiplier. 
The first class constraint $T^a$ can be seen to satisfy the algebra :
\begin{eqnarray}
\{T^a(x),T^b(y)\}_D&=&-gf^{abc}T^c\delta(x-y)\approx 0\\
\{T^a(x),{\cal H}(y)\}_D&=&0
\end{eqnarray}
 The BFV quantization method, in attempting to maintain Lorentz covariance and 
the unitarity of the S-matrix expands the phase space of the theory by making
 the Lagrange multiplier of the theory dynamical, and introducing new (ghost) 
degrees of freedom whose statistics are oppositte to the first class 
constraints of the theory. In our case we will have two pairs of these ghosts 
which are Grassmann fields ;
\begin{eqnarray*}
({\cal C}^a,\bar {\cal P}^a)~~~~;~~~~({\cal P}^a,\bar{\cal C}^a).
\end{eqnarray*}
Therefore, our canonical variables become now
\begin{eqnarray}
Q^A=(A_i^a,\psi,\psi^{\dagger},A_0^a,{\cal C}^a,{\cal P}^a)\\
P^A=(\pi_i^a,\pi_{\psi},\pi_{\psi^{\dagger}},\pi_0^a,\bar{\cal P}^a,
\bar{\cal C}^a)
\end{eqnarray}
Generally, the BFV method introduces the so called complete Hamiltonian [16] 
that enters into the expresion of the generating functional, which is 
 defined as
\begin{eqnarray}
{\cal H}^{comp}={\cal H}_0+\{\Psi,\Omega\}_D
\end{eqnarray}
$\Psi$ is the gauge fermion of the theory and contains all the gauge degrees of
 freedom. $\Omega$ is the BRST charge of the theory, and satisfies :
\begin{eqnarray}
\{\Omega,{\cal H}\}_D&=&0\\
\{\Omega,\Omega\}_D&=&0
\end{eqnarray}
Generally, ${\cal H_0},\Psi$ and $\Omega$ are found as expansions in powers of 
the ghost fields by solving eqs. (83) and (84) above. However, 
in our case, due to the simplicity of the algebra of the constraints, we get 
${\cal H}_0$ to zeroth order, $\Psi$ to first order and $\Omega$ to second 
order in the ghost fields. Thus
\begin{eqnarray}
\Psi&=&\bar{\cal C}^a\chi^a+\bar{\cal P}^aA_0^a\\
\Omega&=&\pi_0^b{\cal P}^b+T^b{\cal C}^b-{1\over 2}\bar{\cal P}_bf^{bcd}
{\cal C}^d{\cal C}^c 
\end{eqnarray}
where $\chi^a$ is a gauge-fixing function
\begin{eqnarray}
\chi_i^a=\partial_iA^{a}_i-f^a(x)
\end{eqnarray}
 The vacuum functional of the theory is given now by the expression
\begin{eqnarray}
Z_0=N\int D\mu(Q,P)\exp i\{\int d^3x(P_A\dot Q^A-{\cal H}^{comp})\}
\end{eqnarray}
where $P_A$ and $Q_A$ are given in eqs.(79) and (80), and 
\begin{eqnarray}
D\mu(Q,P)=DA_i^a DA_0^a D\psi D\bar\psi D\pi_{\psi} D\pi_{\psi^{\dagger}}
D\pi_i^a D\pi_0^a D{\cal C}^a D\bar{\cal C}^a D{\cal P}^a D\bar{\cal P}^a
\nonumber\\
\times\delta(\pi_i^a+{\mu\over 2}\varepsilon_{ij}A^{ja})\delta(\pi_{\psi}-i\psi
^{\dagger})\delta(\pi_{\psi^{\dagger}}\left(Ber||\{\theta_l,\theta_m\}||\right)
^{1\over 2}
\end{eqnarray}
$Ber$ is the superdeterminant, or the Berezinian, which is introduced here due 
to the presence of the fermionic degrees of freedom.
Integrating over the matter and gauge momenta and over $\pi_0^a,\bar{\cal P}^b$
 and ${\cal P}^a$ we get
\begin{eqnarray}
Z_0=N\int DA_{\mu}D\psi D\bar\psi D{\cal C}D\bar{\cal C}\delta(\dot A_0^a(x)-
\partial_iA_i^a(x)+f^a(x))\nonumber\\
\times \exp i\{\int d^3x({\cal L}_{cl}-\bar{\cal C}^a(\partial_\mu D^{\mu ab}
{\cal C}^b))\}
\end{eqnarray}
where
\begin{eqnarray}
{\cal L}_{cl}=i\bar\psi(\partial\!\!\!/-m)\psi-A_{\mu}^aJ^{\mu a}-{\mu\over 2}
A_0^a\varepsilon_{ij}F^{ija}-{\mu\over 2}\varepsilon_{ij}\dot A^{ia}A^{ja}
\nonumber\\
-g{\mu\over 2}f^{abc}A_0^a\varepsilon_{ij}A^{ib}A^{jc}
\end{eqnarray}
and,
\begin{eqnarray}
\bar {\cal C}^a\partial_\mu D^{\mu}_{ ab}{\cal C}^b=\bar{\cal C}^a
(\partial_{\mu}
(\delta_{ab}\partial^{\mu}-f^{acb}A^{\mu}_{c})){\cal C}^b
\end{eqnarray}
The above expression -upon including external sources- coincides with the 
generating functional eq.(59) without the Maxwell term .\\

\section{The S-Matrix Operator} 
$~~~$Although the generating functionals of the theory, eqs.(11),(30) and (59)
  contain all the information of the theory, and can be used to derive the  
scattering amplitudes, it is more convenient to either introduce the path 
integral representation of the S-matrix of the theory, or to construct the 
S-matrix opertator. The latter is particularly convenient for the investigation
  of the imaginary parts of the Greens functions, Feynman diagrams and the 
scattering matrix elements, or generally speaking, for the investigation of the
 unitarity of the theory. We shall first construct the S-matrix operator of the
 pure CSQED, and then generalize the results to the other cases. A peculiar 
property of the pure CSQED is the absence of real topological photons, 
although the propagator and its imaginary part exist (see eqs.(21) and (23) 
for example). As for the operator $\hat A_{\mu}(x) $; we note that 
canonical quantization in covariant gauges allows one to  introduce (as in QED)
  operators for the scalar as well as the longitudinal components of 
$A_{\mu}(x)$, and it can be proven that due to the canonical commutation 
relations, the equation for the propagator
\begin{eqnarray}
D_{\mu\nu}=-i\langle T \hat A_{\mu}(x)\hat A_{\nu}(y)\rangle\nonumber
\end{eqnarray}
coincides with the classical equation (14) and have the same solution, eq.(15).
 However, in the case of pure CS theory, this topological photon does not 
contribute to the physical states of the Hilbert space, which can be  defined
as usual; $\partial_{\mu}\hat A_{\mu}^+|phys\rangle=0$.  
Thus, starting from this result,
 one can unambigously formulate rules for the construction of any
 matrix elements of the different products of this operator. In this sense, 
one can formulate some kind of Wick theorem for the operators of the 
topological CS photon. The S-matrix operator for scalar pure CSQED has been
 construsted in the works [11,13], and those for spinor CSQED in [14]. Here, 
we would like to
elaborate on the construction given in these references.\\
 In pure CSQED, the S-matrix operator formally has the same form as that in 
2+1 dimensional QED,
\begin{eqnarray}
\hat S=T\exp \{iS_{int}(\hat{\bar\psi},\hat\psi,\hat A)\}
\end{eqnarray}
here,
\begin{eqnarray}
S_{int}(\hat{\bar\psi},\hat\psi,\hat A)=\int d^3x :e(\hat{\bar\psi}\gamma^{\mu}
\hat A_{\mu}\hat\psi):
\end{eqnarray}
where ": :" means normal ordering, and $\hat\psi(x)$ and $\hat{\bar\psi}(x)$ 
operators are given as
\begin{eqnarray}
\hat\psi(x)&=&\int {d^3p\over (2\pi)}\sqrt{{m\over E_{\vec p}}}[b
(\vec p)u(p)e^{-ipx}+d^{\dagger}(\vec p)v(p)e^{ipx}]\\
\hat{\bar\psi(x)}&=&\int {d^3p\over (2\pi)}\sqrt{{m\over E_{\vec p}}}[b^
{\dagger}(\vec p)\bar u(p)e^{ipx}+d(\vec p)\bar v(p)e^{-ipx}],
\end{eqnarray}
 $E_{\vec p}=\sqrt{\vec p^2+m^2}$ and $b(\vec p)~(d(\vec p))$ and 
$b^{\dagger}
(\vec p)~(d^{\dagger}(\vec p))$ are respectively the annihilation and creation 
operators of particles (antiparticles) satisfying the usual anticommutation 
relations :
\begin{eqnarray}
[b(\vec p),b^{\dagger}(\vec p')]_+=[d(\vec p),d^{\dagger}(\vec p')]_+=\delta(
\vec p-\vec p')
\end{eqnarray}
 The two-component spinors $u(p), v(p)$ are respectively the 
  positive and negative energy solutions of the free Dirac equation in (2+1)
 dimensions, with the properties:
\begin{eqnarray}
(\hat p-m)u(p)&=&(\hat p+m)v(p)=0\\
\bar u(p)u(p)&=&-\bar v(p)v(p)=1\\
\bar u(p)v(p)&=&\bar v(p)u(p)=0\\
u(p)\bar u(p)&=&{p\!\!\!/+m\over 2m}\\
v(p)\bar v(p)&=&{p\!\!\!/-m\over 2m}
\end{eqnarray}
$~$Let us next our attention to the operator $\hat A_{\mu}$: Using its above 
mentioned properties, we can formulate  
 the following rules of the matrix elements of its 
products : 1)  The vacuum expectation value of the products and the 
T-products
 of only an even number of the operators $A_{\mu}$ is nonvanishing, and reduces
 respectively to the sum of the vacuum expectation values of the product and 
the T-product of two field operators defined as:
\begin{eqnarray}
\langle 0|T(\hat A_{\mu}(x)\hat A_{\nu}(y))|0\rangle&=&-iD_{\mu\nu}(x-y)\\
\langle 0|\hat A_{\mu}(x)\hat A_{\nu}(y)|0\rangle&=&-iD_{\mu\nu}^+
(x-y)\nonumber\\
&=&-i\int {d^3p\over(2\pi)^3}\left[\left({i\over\mu}\varepsilon_{\mu\nu
\lambda}p^{\lambda}-{\alpha\over 2}p_{\mu}{\partial\over \partial^{\nu}}\right)
\delta(p^2)\right]\theta(p_0) e^{ip(x-y)} 
\end{eqnarray}
where $D_{\mu\nu}(x-y)$ is given by eq.(21).
 For example, for four operator product we have:
\begin{eqnarray}
\langle 0|T(\hat A_{\mu}(x)\hat A_{\nu}(y)\hat A_{\lambda}(z)\hat A_{\delta}(u)
|0\rangle&=&(-i)^2\{D_{\mu\nu}(x-y)D_{\lambda\delta}(z-u)+D_{\mu\lambda}(x-z)
\nonumber\\
& &D_{\nu\delta}(y-u)+D_{\mu\delta}(x-u)D_{\nu\lambda}(y-z)\}
\end{eqnarray}
 and so on. 2) All the matrix elements between physical states of the normal 
product
 of any number of the field operators $A_{\mu}$ are equal to zero 
. However, the vacuum expectation value of the product of 
 the normal products of equal number of these operators only is different
from zero. For example:
\begin{eqnarray}
\langle 0|:\hat A_{\mu}(x)\hat A_{\nu}(y)::\hat A_{\lambda}(z)\hat A_{\delta}
(u):|0\rangle=\nonumber\\
(-i)^2\{D^+_{\mu\lambda}(x-z)D^+_{\nu\sigma}(y-u)+D^+_{\mu\sigma}(x-u)D_{\nu
\lambda}^+(y-z)\}
\end{eqnarray} 
 and so on.\\
Thus, the above rules are the same as the Wick rules except that we take into
 account the absence of physical states with free topological photons 
(other than the 
vacuum state !). Therefore, we make now the following observation: All the 
Feynman rules of the theory are identical to those of QED given that one 
replaces 
the Maxwell propagator in internal lines by the CS propagator, and excludes 
diagrams with external photon lines. In mathematical language, the above rules 
 mean that the total set of physical states in the total Hilbert space of the 
theory 
does not contain states with real free topological photons \footnote{The 
absence of the real topological free photons can be seen most generally from 
the fact that the CS term does not contribute to the free classical 
Hamiltonian due to 
its independence of the metric tensor $g_{\mu\nu}$ in curved space-time.} 
, but only the 
physical states of particles and antiparticles. The interesting consequences
 and applications of these statements will be considered in part IV.\\
 Consider now the more general case of MCSQED, where the propagator is given by
 eq.(15) and the free field solutions of the classical equations of motion by
eq.(16). This solution consists of two parts : massive physical part and 
massless topological part. The canonical quantization of the massive part in 
the $\alpha=0$ gauge can be carried out, and gives the following representation
 for the physical massive part  $\hat A_{\mu}^m(x)$ of the operator 
$\hat A_{\mu}\equiv \hat A_{\mu}^m+\hat A_{\mu}^{CS}$\footnote{The details of 
the canonical quantization, which is very similar to the Gupta-Bluer 
quantization will be published in another paper.}
\begin{eqnarray}
\hat A_{\mu}^m(x)={-1\over 2\pi}\int d^3p e^{ipx}\gamma\left({\it e}^
{\delta}_{\mu}(p)-{i\over\mu\gamma}\varepsilon_{\mu\nu\rho}p^{\rho}{\it e}^
{\nu\delta}(p)\right)\delta(p^2-\mu^2\gamma^2)a_{\delta}(p)
\end{eqnarray}
 The S-matrix in this case looks formally the same as (93), but the Wick 
theorem is now the usual one
\begin{eqnarray}
\langle 0|T\hat A_{\mu}(x)\hat A_{\nu}(y)|0\rangle&=&-iD_{\mu\nu}(x-y)+:
\hat A_{\mu}(x)\hat A_{\nu}(y):\\
\langle 0|\hat A_{\mu}(x)\hat A_{\nu}(y)|0\rangle&=&-iD_{\mu\nu}^+(x-y)+:\hat
 A_{\mu}(x)\hat A_{\nu}(y):
\end{eqnarray}
\begin{eqnarray}
\langle 0|:\hat A_{\mu}(x)\hat A_{\nu}(y)::\hat A_{\lambda}(z)\hat A_{\sigma}
(u):|0\rangle=\nonumber\\
(-i)^2\{D_{\mu\lambda}^+(x-z)D_{\nu\sigma}^+(y-u)&+&D_{\mu\sigma}^+(x-u)
D_{\nu\lambda}^+(y-z)\}\nonumber\\
-i\{D_{\mu\lambda}^+(x-z)\langle 0|:\hat A_{\nu}(y)\hat A_{\sigma}(u):|0\rangle
&+&D_{\mu\sigma}^
+(x-u)\langle 0|:\hat A_{\nu}(y)\hat A_{\lambda}(z):|0\rangle\nonumber\\
+D_{\nu\lambda}^+(y-z)\langle 0|:\hat A_{\mu}(x)\hat A_{\sigma}(u):|0\rangle
&+&D_{\nu\sigma}^+(y-w)\langle 0|:\hat A_{\mu}(x)\hat A_{\lambda}(z):|0\rangle
\}\nonumber\\
+\langle 0|:\hat A_{\mu}(x)\hat A_{\nu}(y)\hat A_{\lambda}(z)\hat A_{\sigma}(u)
:|0\rangle& &
\end{eqnarray}
and so on, where $D_{\mu\nu}(x-y)$ is given by eq.(15).Only one important 
exception exists : Any matrix element of the normal product of the operators 
$A_{\mu}$ reduces to that of the normal product of the massive operators 
$A^m_{\mu}$;
\begin{eqnarray}
\langle f|:A_{\mu_1}(x_1)...A_{\mu_n}(x_n):|i\rangle=\langle f|:A_{\mu_1}^m
(x_1)...A_{\mu_n}^m(x_n):|i\rangle
\end{eqnarray}
 Here, $|i\rangle$ and $|j\rangle$ are two arbitrary physical states of the 
total 
Hilbert space of the theory. Now the total set of physical states includes, in 
addition to spinor particles, real massive photons, but never the topological 
massless photons. \\
 The generalization of the S-matrix operator to scalar or spinor pure CSQCD is
straight forward now. For the spinor case, the generating functional is given 
by eq.(90). The S-matrix will have the form
\begin{eqnarray}
S=T\exp\{i\int d^3x\left[-\mu\varepsilon^{\mu\nu\lambda}tr({2i\over 3}e:\hat 
A_{\mu}(x)\hat A_{\nu}(x)\hat A_{\lambda}(x):)-{1\over 2\alpha}tr
(:2e\hat F_{\mu\nu}(x)
[\hat A^{\mu}(x),\hat A^{\nu}(x)]_-\right.\nonumber\\
+e[\hat A_{\mu}(x),\hat A_{\nu}(x)]_-^2:)
\left.+e:\partial^{\mu}\hat{\bar{\cal C}}^a(x)f^{abc}\hat A_{\mu}^b(x)\hat 
{\cal C}^c(x):+e:\hat{\bar\psi}(x)\gamma_{\mu}\hat A^{\mu}(x)\hat\psi(x) :
\right]\}
\end{eqnarray}
The Wick-type theorem for the operators $\hat\psi,\hat{\bar\psi},\hat{\cal C},
\hat{\bar{\cal C}}$ is as usual. 
As for the $\hat A_{\mu}^a$ operator, we have the same rules 
 as in the Abelian case, except that the Greens function will have now an 
additional kronecker delta in the color indices.
\section{Topological Unitarity Identities}
 In this part we are going to investigate the consequences of the peculiar 
property of the CS theories, namely the absence of real topological photons in
 spite of the presence of the propagator and the many-particle Greens function
of the gauge field that contribute to the interaction of the particles quantum
 mechanically ( It is well-known that on the classical level, the CS field do 
not contribute to the interaction of the particles !). We will see that the 
above property of the CS theories leads upon imposing the unitarity condition 
on the theory to very interesting topological unitarity identities. These 
identities have been derived in the work [14]. Here, we essentially follow the 
development in this reference, however, we discuss in more details how do these
 identities hold in the general case when the Maxwell term is present  
 along with the CS term.\\
 We consider first the case of CSQED. The propagator is given by eq.(21) 
, and the S-matrix operator is given by eq.(92). As
we have mentioned above, the absence of the real CS photons means that the 
complete set of physical vector states in the total Hilbert space of the 
theory does 
not contain these topological particles. To investigate the consequences of 
this fact, we introduce the $\hat T$-matrix :
\begin{eqnarray}
\hat S=1-i\hat T
\end{eqnarray}
where $\hat S$ is the S-matrix operator (the energy-momentum conserving 
$\delta$-function has been suppressed). The unitarity of the S-matrix operator
leads to the well-known relation:
\begin{eqnarray}
i\left(\hat T^{\dagger}-\hat T\right)=\hat T\hat T^{\dagger}=2Im\hat T
\end{eqnarray}
 For arbitrary non-diagonal ($|i\rangle\not =|f\rangle$) on-shell matrix 
elements between two physical states of the total Hilbert space, we can write 
the two equivalent relations
\begin{eqnarray}
2Im\langle f|\hat T|i\rangle=\langle f|\hat T\hat T^{\dagger}|i\rangle
\end{eqnarray}
and,
\begin{eqnarray}
2Im\langle f|\hat T|i\rangle=\sum_n\langle f|T|n\rangle\langle n|T^
{\dagger}|i\rangle
\end{eqnarray}
where in eq.(115) we have inserted the complete set of physical states 
{$|n\rangle$} which does not contain the states of the topological photon, but 
only the states of charged particles. From eq.(115) we see that in a given 
order 
of perturbation theory, the Feynman diagrams that contribute to the imaginary 
part on the l.h.s can not have intermediate on-shell topological photon lines
 because {$|n\rangle$} are physical states. On the other hand,however, 
investigating eq.(114) in the framework of perturbation theory, we can see 
that 
diagrams with intermediate on-shell photon lines do appear since the vacuum 
expectation value of the product of the normal products of equal number of 
the operator $A_{\mu}$ (see eq.(105)) does not vanish as a consequence of the 
non-vanishing of the imaginary part of the photon propagator. Therefore, 
demanding the consistency of eqs. (114) and (115) leads to the important 
conclusion that in a given order of perturbation theory, the gauge-invariant 
sum of the imaginary parts of the Feynman diagrams with a given number of 
intermediate on-shell photon lines is equal to zero. The vanishing of this sum
 of the imaginary parts does not mean the vanishing of the sum of the real part
, or the vanishing of the imaginary part of each distinct diagram. As a rule,
the sum of such diagrams will not vanish and will give contribution to the 
process involved. Moreover, each diagram in this sum will be an analytic 
function of invariant variables. The imaginary part of a distinct diagram will
vanish only if the diagram is gauge-invariant. These arguments will be 
demonstrated later when we consider a specific example below.\\
 Now, we illustrate these unitarity identities by an explicit example. Consider
 the case of scattering of a fermion-antifermion pair in one loop order in pure
 CSQED. The S-matrix of this theory is given by eq.(93). 
The gauge-invariant Feynman diagrams with intermediate CS topological photon 
lines are shown in figure 1 below. The analytic expression for the imaginary
part of each of these diagrams is
\begin{eqnarray}
A_a={2g^4\over (2\pi )^3}\int d^3kd^3k'\left(\delta ^+(k^2)\delta ^+(k'^2)
\delta (p+q-k-k')G_{\mu\lambda}(k)G_{\nu\sigma}(k') \right .  \nonumber \\
\times\frac{\bar v(q)\gamma ^\nu (p\!\!\!/-
 k\!\!\!/+m)\gamma ^\mu u(p)\bar u(p')\gamma ^\lambda (p'\!\!\!\!/-k\!\!\!/+m)
\gamma
^\sigma v(q')}{((p-k)^2-m^2+i\epsilon)((p'-k)^2-m^2+i\epsilon)} \left .\right),
\end{eqnarray}
\begin{eqnarray}
A_b={2g^4\over (2\pi )^3}\int d^3kd^3k'\left(\delta ^+(k^2)\delta ^+(k'^2)
\delta
 (p+q-k-k')G_{\mu\lambda}(k)G_{\nu\sigma}(k') \right . \nonumber \\ 
\times  {\bar v(q)\gamma ^\nu (k\!\!\!/- 
q\!\!\!/+m)\gamma ^\mu u(p)\bar u(p')\gamma ^\sigma (p'\!\!\!\!/-k\!\!\!/+m)
\gamma ^\lambda 
v(q')\over ((k-q)^2-m^2+i\epsilon )((p'-k)^2-m^2+i\epsilon)}\left .\right).
\end{eqnarray}
where 
$G_{\mu\nu}(k)=\varepsilon_{\mu\nu\lambda}k^\lambda$, 
and $\delta^+(k^2)=\theta(k_0)\delta(k^2)$. For simplicity, we restrict 
ourselves to the case of forward scattering in which case the imaginary part 
of these diagrams give their contribution to the total cross-section of the 
process. As was shown in [14],a lengthy calculation gives (an overall 
irrelevant multiplicative constant has been suppressed)
\begin{eqnarray}
A_a=-\int d^3k\delta^+(k^2)\left(1+{p.k\over m^2}+{q.k\over p.k}\right)=-A_b
\end{eqnarray}
or,
\begin{eqnarray}
A_a+A_b=0
\end{eqnarray}
 The same result can be obtained in the case of non-forward scattering too. 
This example demonstrates the unitarity identities in the one-loop order.\\
 It is not difficult to generalize the unitarity identities to the case when 
Maxwell-type terms are present. In such cases, one must divide the total gauge
 field propagator in eq.(15) or (62) into two parts (in the $\alpha=0$ gauge
 for 
example ): physical massive part and topological massless part. The operator
$\hat A_{\mu}$ in the exponent of the S-matrix in eq.(12) can be viewed as the 
sum
 of two parts too: the massive physical ($\sim \delta(\gamma^2\mu^2-k^2)$), and
the massless topological part ($\sim\delta(k^2)$). States of the massive photon
 will appear now in the total Hilbert space of the theory. So, imposing the 
unitarity condition on this S-matrix in the sense of eqs.(114) and (115) will 
lead to the appearence of the topological unitarity identities in this case 
too.\\
 For example if we consider the diagrams with two intermediate photon lines 
in the one-loop fermion-antifermion scattering, we get the two unitarity 
identities illustrated diagramatically in figure 2 ( the lines with $\times$
 represent the topological part of the gauge field propagator).
 The first identity means that 
the sum of the four diagrams (which is gauge-invariant) with one on-shell 
intermediate topological photon line is zero. The second identity means the
same for the diagrams with two intermediate on-shell topological lines.\\
 The identities developed above can be also shown to hold outside the 
framework of
 perturbation theory. That they should hold in the non-Abelian case as well, 
could be demonstrated without too much difficulty.  \\
\section{Concluding Remarks}
 In this paper, we have shown that the covariant path integral  
quantization of the theories of scalar and spinor particles interacting 
through the Abelian and non-Abelian pure CS gauge fields, is mathematically 
ill-defined due to the absence of the transverse components of these gauge 
fields.
To define the path integral, it is necessary to introduce into the classical 
action the Maxwell or Maxwell-type (in the non-Abelian theory) term that is 
the only 
bilinear term in the gauge field that does not violate the gauge-invariance of 
the action. This term also guarantees the gauge-invariant regularization and
renormalization of the theory, which becomes then superrenormalizable [2,3].\\
 The generating functionals of the various models considered were constructed,
 and seen to be formally the same as those of QED (or QCD) in 2+1 dimensions, 
with the substitution of the CS gauge field propagator for the photon (or 
gluon) propagator. The CS propagator in these models is seen to consist of two
parts: the first part is the propagator of a real massive photon (gluon) which
contributes to the classical free Hamiltonian, and its states appear in the 
Hilbert 
space of the total set of physical states of the system. The second part is 
that of the 
topological massless photon which does not contribute to the free Hamiltonian,
but leads to additional (in comparison with QED or QCD) interaction between the
 charged particles. The general solution for the free gauge field, when 
constructed in a covariant gauge, was therefore seen to consist of a massive 
part, and a massless topological part.\\
 Taking the limits $\gamma\to\infty$ and $\mu\to 0$ of the propagators and the
general solutions of the gauge fields (see eqs.(21)-(24) ) after 
renormalization,
which is possible due to the finite renormalization of these parameters [5-7],
we get respectively pure CSQED and QED in 2+1 dimensions.\\
 We carried out very carefully the path integral quantization of some models 
with the non-Abelian CS field by the De Witt-Fadeev-Popov and the 
Batalin-Fradkin-Vilkovisky methods, and showed that it is not necessary to 
quantize the dimensionless charge of the theory. First, in the DFP approach we
 use gauge transformations which have zero winding number since the integral 
over the gauge group takes into account only the contributions near the 
identity element of the group (these elements of the group have zero winding 
numbers).Also, the action in the exponent (after path integral quantization) 
is not gauge
 but BRST-invariant, and due to the Grassmann nature of the BRST transformation
, one gets a zero winding number too !. The BFV approach gives the same 
BRST-invariant expression for the action in the exponent of the path integral 
expression.
Finally, the definition of the generating functional (see eqs.(49),(59)) shows 
that for any gauge transformation, the terms proportional to the winding 
number in the path integral expression for the expectation value of any 
observable are 
cancelled due to the normalization of the generating functional (see the 
footnote in the work [3] about the argument of J.Schonfeld in this respect).
 It is well-known that the existence of BRST-invariance in renormalizable gauge
 theories guarantees the implementation of Ward-Fradkin-Takahashi-Slavnov
-Taylor identities, and gauge-invariant renormalization of the theory. This 
invariance, in turn does not require the quantization of the charge.\\
 Unfortunately, a path integral representation of the S-matrix is not 
available for 
theories with the pure CS field. This is because the "in" and "out" limits of
the transverse part of
the pure CS gauge field do not exist. In the general case, when the 
Maxwell-type
 term is included in the action, such a representation can be constructed, and
this will depend only on the "in" and "out" solutions of the massive part of 
the gauge field. We constructed in the general case, the S-matrix 
 operator for all the Abelian and non-Abelian models, and showed that
this operator gives the correct expression for all the Feynman
diagrams of the theory, and formally differs from the usual case of QED and QCD
 in 2+1 dimensions only by a specific type of Wick theorem for the gauge field.
\\
Starting from this S-matrix operator, we have shown that the requirement of 
the unitarity of the S-matrix leads to topolgical unitarity identities that 
were derived in [14]. These identities demand that at
 each order of perturbation theory, the gauge-invariant sum of the imaginary 
parts of the Feynman diagrams with a given number of intermediate on-shell 
topological photon 
lines should vanish. These identities were illustrated by some examples in 
 the Abelian case. The importance of these identities stems
from the fact that they, not only provide additional check of the 
gauge-invariance of the theory, but also highly facilitates the perturbative 
 gauge-invariant calculations of  Feynman diagrams. It is also possible to get
strong restrictions on the dependence on invariant variables of the
 gauge-invariant sum of the real parts of the Feynman diagrams for which the 
gauge-invariant sum of the imaginary parts vanishes (on account of the
 analytic properties of Feynman diagrams in the momentum space representation).

{\bf Acknowledgements:}\\
 We thank I.Tyutin and M.Soloviev from the Lebedev Institute of
 Physics for helpful discussions.\\


\newpage
\begin{thebibliography}{99}
\bibitem{} S.Weinberg in " Understanding the Fundamental Constituents of 
matter" (A.Zichichi, Ed.), Plenum, New York (1978).; A.Linde, Rep.Prog.Phys.
{\bf 42} (1979),389; D.Gross, R.Pisarski and L.Yoffe, Rev.Mod.Phys. {\bf 53}
(1981),43.
\bibitem{} J.Schonfeld, Nucl.Phys.{\bf B 185} (1981), 157.
\bibitem{} S.Deser, R.Jackiw and S.Templeton, Ann.Phys.(N.Y) {\bf 140} (1982),
157; Phys.Rev.Lett.{\bf 48} (1982), 975.
\bibitem{}
C.R.Hagen, Ann.Phys.(N.Y) {\bf 157} (1984), 342.
\bibitem{} G.W.Semenoff, P.Sodano and Y-S.Wu, Phys.Rev.Lett.{\bf 62} (1989),
75; W.Chen, Phys.Lett.{\bf B 251} (1990), 415; D.K.Hong, T.Lee and S.H.Park, 
Phys.Rev.D {\bf 48} (1993), 3918.     
\bibitem{} R.Pisarski and S.Rao, Phys.Rev.D {\bf 32} (1985), 2081; G.Giavarini,
C.P.Martin and F.Ruiz Ruiz, Nucl.Phys.{\bf B 381} (1992), 222.
\bibitem{} S.Coleman and B.Hill, Phys.Lett.{\bf B 159} (1985), 184.
\bibitem{} F.Wilczek and A.Zee, Phys.Rev.Lett.{\bf 51} (1983), 2250.
\bibitem{} F.Wilczek (ed.), " Fractional Statistics and the Superconductivity 
of Anyons ", \\( World Scientific, Singapore 1990); R.Prange and S.Girvin 
(eds.), " The Quantum Hall Effect ", ( Springer, Berlin 1986).
\bibitem{} C.R.Hagen, Phys.Rev.D {\bf 41} (1985), 848;
O.Bergmann and G.Lozano, Ann.Phys.(N.Y) {\bf 229} (1994), 416; O.Bergmann 
and D.Bak, Phys.Rev.D {\bf 51} (1995), 1994.
\bibitem{} 
M.Boz, V.Ya.Fainberg and N.K.Pak, Phys.Lett.{\bf A 207} (1995),1; Ann.Phys.
(N.Y) {\bf 246} (1996), 347. 
\bibitem{} D.Boyanovski, Int.J.Mod.Phys.{\bf A 7} (1992),5917. 
\bibitem{} V.Ya.Fainberg and N.K.Pak, Journ.Theor.Math.Phys.{\bf 103} (1995),
328.
\bibitem{} V.Ya.Fainberg and M.S.Shikakhwa, Phys.Rev.D (1996), (in press).
\bibitem{}
B.S.De Witt, Phys.Rev.{\bf 162} (1967), 1195; L.D.Fadeev and V.N.Popov,
Phys.Lett.{\bf B 25} (1967), 30.
\bibitem{} E.S.Fradkin and G.A.Vilkovisky, Phys.Lett.{\bf B 55} (1975), 224; 
J.A.Batalin and G.A.Vilkovisky, Phys.Lett.{\bf B 69} (1977), 309; E.S.Fradkin 
and T.E.Fradkina, Phys.Lett.{\bf B 72} (1978), 343.
\bibitem{} E.C.Marino, Nucl.Phys.{\bf B 408} (1993),551.
\bibitem{} V.V.Gribov, Nucl.Phys.{\bf B 139} (1978),1.
\bibitem{} C.Becchi, A.Rouet and S.Srora, Ann.Phys.(N.Y) {\bf 98} (1976), 287;
I.V.Tyutin, Lebedev Institute preprint, FIAN (1975) No.39, unpublished.
\bibitem{} J.C.Ward, Phys.Rev.{\bf 77} (1950), 2931; E.S.Fradkin, Sov.Phys.JETP
{\bf 6} (1955), 370; Y.Takahashi, Nuovo C.mm.{\bf 6} (1957), 370; 
A.A.slavnov, Jour.Math.Theor.Phys.{\bf 10} (1972), 99; J.C.Taylor, Nucl.Phys.
{\bf B 33} (1971), 436.
\bibitem{} B.L.voronov and I.V.Tyutin, Theor.Math.Phys.{\bf 50} (1982), 218;
{\bf 52} (1982), 628; B.L.Voronov, P.M.Lavrov and I.V.Tyutin, Sov.J.Nucl.Phys.
{\bf 36} (1982), 292.
\bibitem{} P.A.M.Dirac, " Lectures on Quantum Mechanics", (Yeshiva University,
 New York, 1964).

\end{thebibliography}
\end{document}